\documentstyle[12pt]{article}

\title{Heterogeneous decay of metastable phase on various centers -2 }
\author{V.Kurasov \\
St.Petersburg State University, \\
Department of Mathematical and Computational Physics}

\date{Victor.Kurasov@pobox.spbu.ru }

\begin{document}
\maketitle
\begin{abstract}

A system of a metastable phase with several sorts of
heterogeneous centers is
considered.
An analytical theory for the process of
decay in such a system has been constructed.
The free energy of  formation of a critical embryo  is assumed to be known in
the
 macroscopic approach. At first all
asymptotic cases are investigated and then a general intermediate solution is
suggested.
All approximate transformations are accomplished with the corresponding
numerical
estimates and  analytical justifications.
This is the second part of the manuscript, the first part can be
also found in this archive.

\end{abstract}

\pagebreak

This is the second part of the manuscript, the first part can be
found in the cond-mat e-print archive at the same publishing date.

\section{Intermediate approximate solution for several sorts of centers}

 In the case of
nucleation on one sort of  centers    the expression for $g_i (z)$ is unknown only
when the spectrum is cut off by  exhaustion of  heterogeneous centers.
 So, one doesn't know it in the situation when the converging force of
the operator $S_i$  is very strong. In
the case of the several sorts of centers the situation is
another. One doesn't know every term in $\sum_{j} g_{j}(z)$.
Then
one can come to the situation where according  to the first iteration the spectrum
is cut off by the exhaustion of supersaturation initiated
by droplets formed on some  sort of centers but
in reality  all heterogeneous centers of that sort are exhausted
and those droplets
practically don't consume the vapor  (due to their quantity).
So, it is necessary to have  some more precise expression for
$g_{i}$ which allows to calculate the next iteration for $\theta_i$.

\subsection{Mono-disperse approximation}

The length
corresponding  to the cut-off of the  spectrum by the substance exhaustion
 is practically one and the same for all sorts of the droplets (all
$\Gamma_i$ have approximately one and the same
value\footnote{Alternative
situations correspond to the asymptotic cases and
have been already described.}).
Analyzing the subintergal expression in the equation for  $g_{i}$
in the pseudo homogeneous case
one can see that the
 subintegral expression
 $s= (z-x)^3 \exp(-\Gamma_i \sum_j g_j / \Phi_*)$
connected with a variation of the supersaturation
 in the formula for $g_i$
is  a very sharp function of $x$. It is less than the function
\begin{equation} \label{77}
s_{bel} = \Theta(z-x) (z-x)^3
\end{equation}
and greater than the function
\begin{equation} \label{78}
s_{ab} = \Theta(z-x) (z-x)^3 \exp(-\frac{\Gamma_{i}\sum_j f_{* j}}
{\Phi_{*}} \frac{x^4}{4} )
\ \ .
\end{equation}
Fig. 6 illustrates   the behavior of the values $s_{ab}$ and
$s_{bel}$. It can be seen that they practically
coincide. This is the reason of  applicability of the first iteration
as a good approximation for the precise solution for one sort of centers.

\begin{figure}

\begin{picture}(450,250)
\put(130,113){\vector(1,-1){34}}
\put(207,75){\vector(-1,0){40}}
\put(210,70){$s_{bel}$}
\put(110,108){$s_{ab}$}
\put(60,10){\special{em: moveto}}
\put(290,10){\special{em: lineto}}
\put(290,240){\special{em: lineto}}
\put(60,240){\special{em: lineto}}
\put(60,10){\special{em: lineto}}
\put(65,15){\special{em: moveto}}
\put(285,15){\special{em: lineto}}
\put(285,235){\special{em: lineto}}
\put(65,235){\special{em: lineto}}
\put(65,15){\special{em: lineto}}
\put(100,50){\vector(1,0){150}}
\put(100,50){\vector(0,1){150}}
\put(150,50){\vector(0,1){3}}
\put(200,50){\vector(0,1){3}}
\put(100,100){\vector(1,0){3}}
\put(100,150){\vector(1,0){3}}
\put(90,40){$0$}
\put(148,40){$0.5z$}
\put(198,40){$z$}
\put(82,98){$0.5$}
\put(90,148){$1$}
\put(108,208){$s \ \ (in \ relative \ units ) $}
\put(255,55){$\rho$}
\put(100,50){\special{em: moveto}}
\put(100.50,50.00){\special{em: lineto}}
\put(101.00,50.00){\special{em: lineto}}
\put(101.50,50.00){\special{em: lineto}}
\put(102.00,50.00){\special{em: lineto}}
\put(102.50,50.00){\special{em: lineto}}
\put(103.00,50.00){\special{em: lineto}}
\put(103.50,50.00){\special{em: lineto}}
\put(104.00,50.01){\special{em: lineto}}
\put(104.50,50.01){\special{em: lineto}}
\put(105.00,50.01){\special{em: lineto}}
\put(105.50,50.02){\special{em: lineto}}
\put(106.00,50.02){\special{em: lineto}}
\put(106.50,50.03){\special{em: lineto}}
\put(107.00,50.03){\special{em: lineto}}
\put(107.50,50.04){\special{em: lineto}}
\put(108.00,50.05){\special{em: lineto}}
\put(108.50,50.06){\special{em: lineto}}
\put(109.00,50.07){\special{em: lineto}}
\put(109.50,50.09){\special{em: lineto}}
\put(110.00,50.10){\special{em: lineto}}
\put(110.50,50.12){\special{em: lineto}}
\put(111.00,50.13){\special{em: lineto}}
\put(111.50,50.15){\special{em: lineto}}
\put(112.00,50.17){\special{em: lineto}}
\put(112.50,50.20){\special{em: lineto}}
\put(113.00,50.22){\special{em: lineto}}
\put(113.50,50.25){\special{em: lineto}}
\put(114.00,50.27){\special{em: lineto}}
\put(114.50,50.30){\special{em: lineto}}
\put(115.00,50.34){\special{em: lineto}}
\put(115.50,50.37){\special{em: lineto}}
\put(116.00,50.41){\special{em: lineto}}
\put(116.50,50.45){\special{em: lineto}}
\put(117.00,50.49){\special{em: lineto}}
\put(117.50,50.54){\special{em: lineto}}
\put(118.00,50.58){\special{em: lineto}}
\put(118.50,50.63){\special{em: lineto}}
\put(119.00,50.69){\special{em: lineto}}
\put(119.50,50.74){\special{em: lineto}}
\put(120.00,50.80){\special{em: lineto}}
\put(120.50,50.86){\special{em: lineto}}
\put(121.00,50.93){\special{em: lineto}}
\put(121.50,50.99){\special{em: lineto}}
\put(122.00,51.06){\special{em: lineto}}
\put(122.50,51.14){\special{em: lineto}}
\put(123.00,51.22){\special{em: lineto}}
\put(123.50,51.30){\special{em: lineto}}
\put(124.00,51.38){\special{em: lineto}}
\put(124.50,51.47){\special{em: lineto}}
\put(125.00,51.56){\special{em: lineto}}
\put(125.50,51.66){\special{em: lineto}}
\put(126.00,51.76){\special{em: lineto}}
\put(126.50,51.86){\special{em: lineto}}
\put(127.00,51.97){\special{em: lineto}}
\put(127.50,52.08){\special{em: lineto}}
\put(128.00,52.20){\special{em: lineto}}
\put(128.50,52.31){\special{em: lineto}}
\put(129.00,52.44){\special{em: lineto}}
\put(129.50,52.57){\special{em: lineto}}
\put(130.00,52.70){\special{em: lineto}}
\put(130.50,52.84){\special{em: lineto}}
\put(131.00,52.98){\special{em: lineto}}
\put(131.50,53.13){\special{em: lineto}}
\put(132.00,53.28){\special{em: lineto}}
\put(132.50,53.43){\special{em: lineto}}
\put(133.00,53.59){\special{em: lineto}}
\put(133.50,53.76){\special{em: lineto}}
\put(134.00,53.93){\special{em: lineto}}
\put(134.50,54.11){\special{em: lineto}}
\put(135.00,54.29){\special{em: lineto}}
\put(135.50,54.47){\special{em: lineto}}
\put(136.00,54.67){\special{em: lineto}}
\put(136.50,54.86){\special{em: lineto}}
\put(137.00,55.07){\special{em: lineto}}
\put(137.50,55.27){\special{em: lineto}}
\put(138.00,55.49){\special{em: lineto}}
\put(138.50,55.71){\special{em: lineto}}
\put(139.00,55.93){\special{em: lineto}}
\put(139.50,56.16){\special{em: lineto}}
\put(140.00,56.40){\special{em: lineto}}
\put(140.50,56.64){\special{em: lineto}}
\put(141.00,56.89){\special{em: lineto}}
\put(141.50,57.15){\special{em: lineto}}
\put(142.00,57.41){\special{em: lineto}}
\put(142.50,57.68){\special{em: lineto}}
\put(143.00,57.95){\special{em: lineto}}
\put(143.50,58.23){\special{em: lineto}}
\put(144.00,58.52){\special{em: lineto}}
\put(144.50,58.81){\special{em: lineto}}
\put(145.00,59.11){\special{em: lineto}}
\put(145.50,59.42){\special{em: lineto}}
\put(146.00,59.73){\special{em: lineto}}
\put(146.50,60.05){\special{em: lineto}}
\put(147.00,60.38){\special{em: lineto}}
\put(147.50,60.72){\special{em: lineto}}
\put(148.00,61.06){\special{em: lineto}}
\put(148.50,61.41){\special{em: lineto}}
\put(149.00,61.76){\special{em: lineto}}
\put(149.50,62.13){\special{em: lineto}}
\put(150.00,62.50){\special{em: lineto}}
\put(150.50,62.88){\special{em: lineto}}
\put(151.00,63.27){\special{em: lineto}}
\put(151.50,63.66){\special{em: lineto}}
\put(152.00,64.06){\special{em: lineto}}
\put(152.50,64.47){\special{em: lineto}}
\put(153.00,64.89){\special{em: lineto}}
\put(153.50,65.31){\special{em: lineto}}
\put(154.00,65.75){\special{em: lineto}}
\put(154.50,66.19){\special{em: lineto}}
\put(155.00,66.64){\special{em: lineto}}
\put(155.50,67.10){\special{em: lineto}}
\put(156.00,67.56){\special{em: lineto}}
\put(156.50,68.04){\special{em: lineto}}
\put(157.00,68.52){\special{em: lineto}}
\put(157.50,69.01){\special{em: lineto}}
\put(158.00,69.51){\special{em: lineto}}
\put(158.50,70.02){\special{em: lineto}}
\put(159.00,70.54){\special{em: lineto}}
\put(159.50,71.06){\special{em: lineto}}
\put(160.00,71.60){\special{em: lineto}}
\put(160.50,72.14){\special{em: lineto}}
\put(161.00,72.70){\special{em: lineto}}
\put(161.50,73.26){\special{em: lineto}}
\put(162.00,73.83){\special{em: lineto}}
\put(162.50,74.41){\special{em: lineto}}
\put(163.00,75.00){\special{em: lineto}}
\put(163.50,75.60){\special{em: lineto}}
\put(164.00,76.21){\special{em: lineto}}
\put(164.50,76.83){\special{em: lineto}}
\put(165.00,77.46){\special{em: lineto}}
\put(165.50,78.10){\special{em: lineto}}
\put(166.00,78.75){\special{em: lineto}}
\put(166.50,79.41){\special{em: lineto}}
\put(167.00,80.08){\special{em: lineto}}
\put(167.50,80.75){\special{em: lineto}}
\put(168.00,81.44){\special{em: lineto}}
\put(168.50,82.14){\special{em: lineto}}
\put(169.00,82.85){\special{em: lineto}}
\put(169.50,83.57){\special{em: lineto}}
\put(170.00,84.30){\special{em: lineto}}
\put(170.50,85.04){\special{em: lineto}}
\put(171.00,85.79){\special{em: lineto}}
\put(171.50,86.55){\special{em: lineto}}
\put(172.00,87.32){\special{em: lineto}}
\put(172.50,88.11){\special{em: lineto}}
\put(173.00,88.90){\special{em: lineto}}
\put(173.50,89.71){\special{em: lineto}}
\put(174.00,90.52){\special{em: lineto}}
\put(174.50,91.35){\special{em: lineto}}
\put(175.00,92.19){\special{em: lineto}}
\put(175.50,93.04){\special{em: lineto}}
\put(176.00,93.90){\special{em: lineto}}
\put(176.50,94.77){\special{em: lineto}}
\put(177.00,95.65){\special{em: lineto}}
\put(177.50,96.55){\special{em: lineto}}
\put(178.00,97.46){\special{em: lineto}}
\put(178.50,98.37){\special{em: lineto}}
\put(179.00,99.30){\special{em: lineto}}
\put(179.50,100.25){\special{em: lineto}}
\put(180.00,101.20){\special{em: lineto}}
\put(180.50,102.17){\special{em: lineto}}
\put(181.00,103.14){\special{em: lineto}}
\put(181.50,104.13){\special{em: lineto}}
\put(182.00,105.14){\special{em: lineto}}
\put(182.50,106.15){\special{em: lineto}}
\put(183.00,107.18){\special{em: lineto}}
\put(183.50,108.22){\special{em: lineto}}
\put(184.00,109.27){\special{em: lineto}}
\put(184.50,110.34){\special{em: lineto}}
\put(185.00,111.41){\special{em: lineto}}
\put(185.50,112.50){\special{em: lineto}}
\put(186.00,113.61){\special{em: lineto}}
\put(186.50,114.72){\special{em: lineto}}
\put(187.00,115.85){\special{em: lineto}}
\put(187.50,116.99){\special{em: lineto}}
\put(188.00,118.15){\special{em: lineto}}
\put(188.50,119.32){\special{em: lineto}}
\put(189.00,120.50){\special{em: lineto}}
\put(189.50,121.69){\special{em: lineto}}
\put(190.00,122.90){\special{em: lineto}}
\put(190.50,124.12){\special{em: lineto}}
\put(191.00,125.36){\special{em: lineto}}
\put(191.50,126.61){\special{em: lineto}}
\put(192.00,127.87){\special{em: lineto}}
\put(192.50,129.15){\special{em: lineto}}
\put(193.00,130.44){\special{em: lineto}}
\put(193.50,131.74){\special{em: lineto}}
\put(194.00,133.06){\special{em: lineto}}
\put(194.50,134.39){\special{em: lineto}}
\put(195.00,135.74){\special{em: lineto}}
\put(195.50,137.10){\special{em: lineto}}
\put(196.00,138.47){\special{em: lineto}}
\put(196.50,139.86){\special{em: lineto}}
\put(197.00,141.27){\special{em: lineto}}
\put(197.50,142.69){\special{em: lineto}}
\put(198.00,144.12){\special{em: lineto}}
\put(198.50,145.57){\special{em: lineto}}
\put(199.00,147.03){\special{em: lineto}}
\put(199.50,148.51){\special{em: lineto}}
\put(200.00,150.00){\special{em: lineto}}
\put(100,50){\special{em: moveto}}
\put(100.50,50.00){\special{em: lineto}}
\put(101.00,50.00){\special{em: lineto}}
\put(101.50,50.00){\special{em: lineto}}
\put(102.00,50.00){\special{em: lineto}}
\put(102.50,50.00){\special{em: lineto}}
\put(103.00,50.00){\special{em: lineto}}
\put(103.50,50.00){\special{em: lineto}}
\put(104.00,50.00){\special{em: lineto}}
\put(104.50,50.00){\special{em: lineto}}
\put(105.00,50.00){\special{em: lineto}}
\put(105.50,50.00){\special{em: lineto}}
\put(106.00,50.00){\special{em: lineto}}
\put(106.50,50.00){\special{em: lineto}}
\put(107.00,50.00){\special{em: lineto}}
\put(107.50,50.00){\special{em: lineto}}
\put(108.00,50.00){\special{em: lineto}}
\put(108.50,50.00){\special{em: lineto}}
\put(109.00,50.00){\special{em: lineto}}
\put(109.50,50.00){\special{em: lineto}}
\put(110.00,50.00){\special{em: lineto}}
\put(110.50,50.00){\special{em: lineto}}
\put(111.00,50.00){\special{em: lineto}}
\put(111.50,50.00){\special{em: lineto}}
\put(112.00,50.00){\special{em: lineto}}
\put(112.50,50.00){\special{em: lineto}}
\put(113.00,50.00){\special{em: lineto}}
\put(113.50,50.00){\special{em: lineto}}
\put(114.00,50.00){\special{em: lineto}}
\put(114.50,50.00){\special{em: lineto}}
\put(115.00,50.00){\special{em: lineto}}
\put(115.50,50.00){\special{em: lineto}}
\put(116.00,50.00){\special{em: lineto}}
\put(116.50,50.00){\special{em: lineto}}
\put(117.00,50.00){\special{em: lineto}}
\put(117.50,50.00){\special{em: lineto}}
\put(118.00,50.00){\special{em: lineto}}
\put(118.50,50.00){\special{em: lineto}}
\put(119.00,50.00){\special{em: lineto}}
\put(119.50,50.00){\special{em: lineto}}
\put(120.00,50.00){\special{em: lineto}}
\put(120.50,50.00){\special{em: lineto}}
\put(121.00,50.00){\special{em: lineto}}
\put(121.50,50.00){\special{em: lineto}}
\put(122.00,50.00){\special{em: lineto}}
\put(122.50,50.00){\special{em: lineto}}
\put(123.00,50.00){\special{em: lineto}}
\put(123.50,50.01){\special{em: lineto}}
\put(124.00,50.01){\special{em: lineto}}
\put(124.50,50.01){\special{em: lineto}}
\put(125.00,50.01){\special{em: lineto}}
\put(125.50,50.01){\special{em: lineto}}
\put(126.00,50.01){\special{em: lineto}}
\put(126.50,50.02){\special{em: lineto}}
\put(127.00,50.02){\special{em: lineto}}
\put(127.50,50.03){\special{em: lineto}}
\put(128.00,50.03){\special{em: lineto}}
\put(128.50,50.04){\special{em: lineto}}
\put(129.00,50.04){\special{em: lineto}}
\put(129.50,50.05){\special{em: lineto}}
\put(130.00,50.06){\special{em: lineto}}
\put(130.50,50.07){\special{em: lineto}}
\put(131.00,50.08){\special{em: lineto}}
\put(131.50,50.09){\special{em: lineto}}
\put(132.00,50.11){\special{em: lineto}}
\put(132.50,50.12){\special{em: lineto}}
\put(133.00,50.14){\special{em: lineto}}
\put(133.50,50.16){\special{em: lineto}}
\put(134.00,50.19){\special{em: lineto}}
\put(134.50,50.22){\special{em: lineto}}
\put(135.00,50.25){\special{em: lineto}}
\put(135.50,50.28){\special{em: lineto}}
\put(136.00,50.32){\special{em: lineto}}
\put(136.50,50.36){\special{em: lineto}}
\put(137.00,50.41){\special{em: lineto}}
\put(137.50,50.46){\special{em: lineto}}
\put(138.00,50.52){\special{em: lineto}}
\put(138.50,50.58){\special{em: lineto}}
\put(139.00,50.65){\special{em: lineto}}
\put(139.50,50.72){\special{em: lineto}}
\put(140.00,50.80){\special{em: lineto}}
\put(140.50,50.89){\special{em: lineto}}
\put(141.00,50.99){\special{em: lineto}}
\put(141.50,51.10){\special{em: lineto}}
\put(142.00,51.21){\special{em: lineto}}
\put(142.50,51.34){\special{em: lineto}}
\put(143.00,51.47){\special{em: lineto}}
\put(143.50,51.61){\special{em: lineto}}
\put(144.00,51.77){\special{em: lineto}}
\put(144.50,51.93){\special{em: lineto}}
\put(145.00,52.11){\special{em: lineto}}
\put(145.50,52.30){\special{em: lineto}}
\put(146.00,52.50){\special{em: lineto}}
\put(146.50,52.71){\special{em: lineto}}
\put(147.00,52.94){\special{em: lineto}}
\put(147.50,53.18){\special{em: lineto}}
\put(148.00,53.43){\special{em: lineto}}
\put(148.50,53.70){\special{em: lineto}}
\put(149.00,53.99){\special{em: lineto}}
\put(149.50,54.28){\special{em: lineto}}
\put(150.00,54.60){\special{em: lineto}}
\put(150.50,54.93){\special{em: lineto}}
\put(151.00,55.27){\special{em: lineto}}
\put(151.50,55.64){\special{em: lineto}}
\put(152.00,56.01){\special{em: lineto}}
\put(152.50,56.41){\special{em: lineto}}
\put(153.00,56.82){\special{em: lineto}}
\put(153.50,57.25){\special{em: lineto}}
\put(154.00,57.69){\special{em: lineto}}
\put(154.50,58.15){\special{em: lineto}}
\put(155.00,58.63){\special{em: lineto}}
\put(155.50,59.13){\special{em: lineto}}
\put(156.00,59.64){\special{em: lineto}}
\put(156.50,60.17){\special{em: lineto}}
\put(157.00,60.72){\special{em: lineto}}
\put(157.50,61.28){\special{em: lineto}}
\put(158.00,61.86){\special{em: lineto}}
\put(158.50,62.46){\special{em: lineto}}
\put(159.00,63.07){\special{em: lineto}}
\put(159.50,63.70){\special{em: lineto}}
\put(160.00,64.34){\special{em: lineto}}
\put(160.50,65.00){\special{em: lineto}}
\put(161.00,65.68){\special{em: lineto}}
\put(161.50,66.37){\special{em: lineto}}
\put(162.00,67.07){\special{em: lineto}}
\put(162.50,67.79){\special{em: lineto}}
\put(163.00,68.53){\special{em: lineto}}
\put(163.50,69.27){\special{em: lineto}}
\put(164.00,70.04){\special{em: lineto}}
\put(164.50,70.81){\special{em: lineto}}
\put(165.00,71.60){\special{em: lineto}}
\put(165.50,72.40){\special{em: lineto}}
\put(166.00,73.22){\special{em: lineto}}
\put(166.50,74.04){\special{em: lineto}}
\put(167.00,74.88){\special{em: lineto}}
\put(167.50,75.73){\special{em: lineto}}
\put(168.00,76.59){\special{em: lineto}}
\put(168.50,77.46){\special{em: lineto}}
\put(169.00,78.34){\special{em: lineto}}
\put(169.50,79.23){\special{em: lineto}}
\put(170.00,80.13){\special{em: lineto}}
\put(170.50,81.04){\special{em: lineto}}
\put(171.00,81.96){\special{em: lineto}}
\put(171.50,82.89){\special{em: lineto}}
\put(172.00,83.83){\special{em: lineto}}
\put(172.50,84.78){\special{em: lineto}}
\put(173.00,85.73){\special{em: lineto}}
\put(173.50,86.69){\special{em: lineto}}
\put(174.00,87.67){\special{em: lineto}}
\put(174.50,88.64){\special{em: lineto}}
\put(175.00,89.63){\special{em: lineto}}
\put(175.50,90.63){\special{em: lineto}}
\put(176.00,91.63){\special{em: lineto}}
\put(176.50,92.64){\special{em: lineto}}
\put(177.00,93.65){\special{em: lineto}}
\put(177.50,94.68){\special{em: lineto}}
\put(178.00,95.71){\special{em: lineto}}
\put(178.50,96.75){\special{em: lineto}}
\put(179.00,97.79){\special{em: lineto}}
\put(179.50,98.85){\special{em: lineto}}
\put(180.00,99.91){\special{em: lineto}}
\put(180.50,100.97){\special{em: lineto}}
\put(181.00,102.05){\special{em: lineto}}
\put(181.50,103.13){\special{em: lineto}}
\put(182.00,104.22){\special{em: lineto}}
\put(182.50,105.32){\special{em: lineto}}
\put(183.00,106.42){\special{em: lineto}}
\put(183.50,107.53){\special{em: lineto}}
\put(184.00,108.65){\special{em: lineto}}
\put(184.50,109.78){\special{em: lineto}}
\put(185.00,110.92){\special{em: lineto}}
\put(185.50,112.06){\special{em: lineto}}
\put(186.00,113.22){\special{em: lineto}}
\put(186.50,114.38){\special{em: lineto}}
\put(187.00,115.55){\special{em: lineto}}
\put(187.50,116.73){\special{em: lineto}}
\put(188.00,117.92){\special{em: lineto}}
\put(188.50,119.12){\special{em: lineto}}
\put(189.00,120.33){\special{em: lineto}}
\put(189.50,121.55){\special{em: lineto}}
\put(190.00,122.78){\special{em: lineto}}
\put(190.50,124.03){\special{em: lineto}}
\put(191.00,125.28){\special{em: lineto}}
\put(191.50,126.54){\special{em: lineto}}
\put(192.00,127.82){\special{em: lineto}}
\put(192.50,129.11){\special{em: lineto}}
\put(193.00,130.40){\special{em: lineto}}
\put(193.50,131.72){\special{em: lineto}}
\put(194.00,133.04){\special{em: lineto}}
\put(194.50,134.38){\special{em: lineto}}
\put(195.00,135.73){\special{em: lineto}}
\put(195.50,137.09){\special{em: lineto}}
\put(196.00,138.47){\special{em: lineto}}
\put(196.50,139.86){\special{em: lineto}}
\put(197.00,141.27){\special{em: lineto}}
\put(197.50,142.69){\special{em: lineto}}
\put(198.00,144.12){\special{em: lineto}}
\put(198.50,145.57){\special{em: lineto}}
\put(199.00,147.03){\special{em: lineto}}
\put(199.50,148.51){\special{em: lineto}}
\put(200.00,150.00){\special{em: lineto}}
\put(200.00,150.00){\special{em: moveto}}
\put(200.00,150.00){\special{em: lineto}}
\put(200,50){\special{em: moveto}}
\put(200,150){\special{em: lineto}}
\end{picture}

{\small \it
\begin{center}
Figure 6.
\\
Functions $s_{ab} $ and $s_{bel}$.
   \end{center}
}

\end{figure}

Let us introduce an approximation for this function. One has to
extract the region of the sizes of the droplets  which are essential in
the vapor consumption.
This consumption  is essential when
\begin{equation} \label{79}
z
\approx
\Delta x
\end{equation}
where $\Delta x$ is the length of the cut-off by the supersaturation.
Certainly,
the sizes of the droplets  which are essential in
the vapor consumption must be  smaller than
$\Delta x$ because the effectiveness of the
 iteration procedure in the homogeneous  case
is based on the fact
that the droplets formed at the almost ideal supersaturation
determine  the formation of the spectrum.
One can see  that $s$ is a very sharp function of $x$ and this is the
reason for the monodisperce approximaion for $s$.
Now one has to define  characteristics  of this approximation.

The condition for the differential width
is that
the subintegral function falls two times in comparison with an amplitude
value.
For the differential half-width $\delta_{1/2} x$
one can get  the following expression
\begin{equation} \label{80}
\delta_{1/2}x = ( 1 - \frac{1}{2^{1/3}} ) x \ \ .
\end{equation}
The integral half-width $\Delta_{1/2}x$ can be obtained from the
corresponding equation
\begin{equation} \label{81}
N_{ess}x^3 = f_{*\ i}\frac{x^4}{4} n_{\infty} \ \ ,
\end{equation}
where $N_{ess} $ is the characteristic number of the  droplets obtained as
$N_{ess}=f_{*\ i}\Delta_{1/2}x n_{\infty}$ and the r.h.s. of the previous
equation is the first iteration approximation. This   gives
\begin{equation} \label{82}
\Delta_{1/2}x = \frac{1}{4} x
\end{equation}
and it practically coincides with $\delta_{1/2}x$.

The subintegral function $s$ is now decomposed into
the essential part where
$$x \leq \frac{z}{4}$$ and the tail where $$x \geq
 \frac{z}{4}$$
The tail will be neglected and  due to the small  relative size of the essential
part
 the mono-disperse approximation
 of sizes for the droplets  in the essential part will be used.
As the result one can  get an approximation
\begin{equation} \label{83}
g(z) = \frac{N(z/4)}{n_{\infty}}z^3 \ \ ,
\end{equation}
where $N(z/4)$ is the number of the droplets appeared from $x=0$ till $x=z/4$.

Since
the spectrum is  cut off by the exhaustion of the supersaturation  in
a frontal (sharp) manner,
 the value of $g_i$ is unessential
before $z=\Delta x$
because it is  small. After the
moment of the cut-off
it is unessential also because there is no formation of the droplets.
So,
instead of the previous approximation one can use
\begin{equation} \label{84}
g_{i}(z) = \frac{N_{i}(\Delta_{i} x/4)}{n_{\infty}}z^3 \ \ .
\end{equation}

Now let us turn to the heterogeneous case.
The exhaustion of
 heterogeneous centers makes the subintegral function
more sharp and the mono-disperse approximation  becomes  at
$x \sim \Delta_{i}x$ even better than in the pseudo homogeneous situation.
 Certainly, one has to use $ N(\Delta_{i} x /4)$
calculated with account of
 exhaustion of  heterogeneous centers (but at the coordinate, obtained without
any account of  exhaustion of  heterogeneous centers).

Fig. 7 illustrates the form of the spectrum
 on the base  of "mono-disperse approximation".
The case of the pseudo homogeneous condensation is considered.
Three curves are drawn: the spectrum in the mono-disperse approximation
$f_{appr}$   ;
 the spectrum in the first iteration $f_{1}$ which can
be considered as a very precise approximation and the numerical solution $f$.
Two of them $f$ and $f_1$ coincide and correspond to a thick line.
The back side
in the first iteration is more sharp  than  in the "mono-disperse
approximation".  Nevertheless the deviation isn't so essential.
It can be eliminated by a perturbation theory.

\begin{figure}



{\small \it
\begin{center}
Figure 7.
\\
Functions $f$, $f_1$ and $f_{appr}$.
   \end{center}
}

\end{figure}

One can  get $ N(\Delta_{i}x/4)$
from the solution of the equations for the
separate condensation process  because
only the lowest length of the cut-off is nesessary. This length is given without
any cross influence
taking into account due to the frontal character of the back side of
the spectrum. Then one has to repeat this procedure with the already known
$N_i$.

\subsection{Final iterations}

Now one can see a way to solve the system of the condensation equations.

{\bf A. Method of separate lengths. }

At first one has to solve  equations for all separate nucleation processes
\begin{equation} \label{85}
g_{i} = f_{*\ i}  \int_{0}^{z} (z-x)^{3}
\exp ( -\Gamma_{i} \frac{ g_{i}  }
{ \Phi_{*} } )
\theta_{i} dx
\equiv
G_{i}(g_{i}, \theta_{i} ) \ \ ,
\end{equation}
\begin{equation} \label{86}
\theta_{i} = \exp ( - f_{*\ i} \frac{n_{\infty}}{\eta_{tot\ i}} \int_{0}^{z}
\exp ( - \Gamma_{i} \frac{ g_{i}  }
{ \Phi_{*} } ) dx )
\equiv
S_i( g_{i})
\end{equation}
for every $i$.
This solution is given by the standard iteration procedure.
In the second approximation
one can calculate the value of $\theta_{(2)}(\Delta_{i}x/4)$
explicitly
\begin{equation} \label{87}
\theta_{i (2)}(\Delta_{i}x/4) =
\exp(- f_{*\ i} \frac{n_{\infty}}
{\eta_{tot\ i}} \int_{0}^{\Delta_{i}x/4}
\exp(-\frac{\Gamma_{i}f_{*\ i}}{4 \Phi_{*}} z^4) dx )
=
\exp(- f_{*\ i} \frac{n_{\infty}}{\eta_{tot\ i}}
(\frac{\Gamma_{i}f_{*\ i}}{4 \Phi_{*}})^{-1/4} C )    \ ,
\end{equation}
where
\begin{equation} \label{89}
C= \int_{0}^{1/4} \exp(-z^4) dx \approx 0.25 \ \ .
\end{equation}
Then
\begin{equation} \label{90}
N_{i (2)}(\Delta_{i}x/4) = \eta_{tot\ i}(1-\theta_{i (2)}(\Delta_{i}x/4)) \ \ .
\end{equation}
One has to fulfil
these calculations for every sort of the heterogeneous centers.
Considering these approximations as the initial ones it is necessary to
do only one more step of the iteration procedure to get the suitable results.
They will be marked as the 'final' ones.
One has to
calculate $\theta_{i\ final}(\infty)$
and $N_{final\ i}(\infty)$
\begin{equation} \label{91}
\theta_{i\ final}(z) =
\exp[- f_{*\ i} \frac{n_{\infty}}{\eta_{tot\ i}} \int_{0}^{z}
\exp(-
\frac{\sum_{j}\Gamma_{i}N_{j\ (2)}(\Delta_{j}x/4)}
{ n_{\infty}\Phi_{*}} x^3) dx]
\ \ ,
\end{equation}
\begin{equation} \label{92}
N_{final\ i}(\infty) = \eta_{tot\ i}[
1-
\exp(- f_{*\ i} \frac{n_{\infty}}{\eta_{tot\ i}}
(\frac{\sum_{j}\Gamma_{i}N_{j\ (2)}(\Delta_{j}x/4)}
{ n_{\infty}\Phi_{*}})^{-1/3} B
]
\ \ .
\end{equation}

These expressions
are valid under the reasonable separation of  heterogeneous
centers into  sorts when the centers with  approximately one and the same height of
the activation barrier are considered as one sort. Meanwhile it is obvious
that when one splits one sort into very many sub-sorts then one can formally
attain a
wrong result.

Let us suppose that the number of  heterogeneous centers
ensures
$h_i \gg 1$,
i.e. the fall of the supersaturation leads to the interruption  of
 formation of
the droplets.
One can split this sort into so many sub-sorts
$(j)$ that for every sub-sort
$h_{ij} \ll 1$,
i.e. one can see the exhaustion of the sub-sort in the separate process
of the condensation.
Moreover, one can assume that this exhaustion is finished before
$z$ attains the quarter
of the length of the spectrum initiated by the fall of the supersaturation.
Since $\Delta_{ij} x$
is proportional to $\eta_{ij \ tot}^{-1/4}$
(where $\eta_{ij \ tot}$
is the total number of the centers of the given sub-sort) and $\delta_{ij} x $
doesn't depend on this quantity,  the required property is obvious.
Then after the summation over all sub-sorts one can  see that
the total number of the droplets formed up to $\Delta_i x / 4$
coincides with the total number of the heterogeneous centers.
This conclusion is wrong.

The reason of the error is that the width of the spectrum is
smaller than the width in the process of the separate formation.
Evidently, the mono-disperse
approximation doesn't work at such distances.
Under the reasonable definition
of the heterogeneous sorts all characteristic lengths are different.
So, ordinary, there is no such effect in this situation.
But still one has to correct these results. The simplest way is to consider
all centers with the difference in the activation barrier height less than one thermal unit
as one and the same sort.
This solves the problem, but this requirement is rather artificial.

Now  these results  will be corrected in more elegant way.
The mono-disperse approximation
leads to some already defined functional form for $g$
and for $\zeta$. The form is already known and now  only
 parameters in these functional dependencies have to be determined.

{\bf B. Method of the unique length. }

The unique length of the spectrum $\Delta x $ will be chosen.
At this very moment  $\Delta x $ is unknown, but
satisfies the following inequality:
$$
\Delta x  \leq \Delta_i x
$$
for every sort of the heterogeneous centers. This leads to
\begin{equation}
\label{ggg}
          g_i(z) = \frac{N_i(\Delta x  / 4)}{n_{\infty}} z^3
\ \ ,
\end{equation}
which gives
\begin{equation}
N_{i(2)}(\Delta x / 4) = \eta_{i\ tot}(1 -  \theta_{i(2)}(\Delta x / 4))
\ \ ,
\end{equation}
\begin{equation}
\label{ff}
         \theta_{i\ (2)} (\Delta x / 4)=
\exp[-f_{*\ i}
\frac{n_{\infty}}{\eta_{tot\ i}}
\int_0^z \exp(-\frac{\Gamma x^3}
{\Phi_* n_{\infty}} \sum_j N_j(\Delta x / 4)) dx ]
\ \ .
\end{equation}
The weak dependence $\Gamma_i \approx \Gamma$
on the sort of the centers is neglected here for
simplicity, while the strong dependence $f_{* i}$
on the sort of centers
is taken into account. After the substitution one can  get the system
of the algebraic equations for $N_i(\Delta x / 4)$
\begin{equation}
\label{ghj}
          N_i (\Delta x / 4) =
\eta_{i\ tot}
(1 - \exp(-f_{*\ i}
\frac{n_{\infty}}{\eta_{tot\ i}}
\int_0^{\Delta x / 4}
\exp(-
\frac{\Gamma}{\Phi_*}
\frac{x^3}{n_{\infty}}
\sum_j N_j(\Delta x
/ 4)) dx))                 \ \ .
\end{equation}

Let us simplify the last system. To calculate the integral one can note
that
\begin{equation}
                \int_0^x \exp(-x^3) dx \approx x \ \ \ \ x \leq 1/4
                \ \ .
\end{equation}
So, the trivial dependence of the r.h.s. on $N_j$
disappears. One can come  to
        \begin{equation}
\label{sf}
         N_i(\Delta x / 4)  = \eta_{i\ tot} ( 1 - \exp(-f_{*\ i}
\frac{n_{\infty}}{\eta_{tot\ i}}
\frac{\Delta
x }{4}))
\ \ .
\end{equation}

On the other hand one can use the sense of $\Delta x $
as the half-width of the spectrum due to the fall of the supersaturation:
\begin{equation}
\frac{\Delta x^3}{n_{\infty}} \sum_j
         N_i(\Delta x / 4)  \frac{\Gamma}{\Phi_*}
= 1
\ \ .
\end{equation}

After the substitution one can get the equation for
$\Delta x$
\begin{equation}
\frac{\Delta x^3}{n_{\infty}} \sum_j \eta_{j\ tot}
(1 - \exp( - f_{*\ j}
\frac{n_{\infty}}{\eta_{tot\ j}}
\frac{\Delta x}{4})) \frac{\Gamma}{\Phi_*}
= 1
 \ \ .
\end{equation}

To solve this
algebraic equation one can choose the initial approximation
as a solution of the following equation
\begin{equation}
\frac{\Delta x^4}{4 } \sum_j
  f_{*\ j}
 \frac{\Gamma}{\Phi_*}
= 1
\ \ .
\end{equation}
Iterations are constructed according to
\begin{equation}
\frac{\Delta_{(i)} x^3}{n_{\infty}} \sum_j \eta_{j\ tot}
(1 - \exp( - f_{*\ j}
\frac{n_{\infty}}{\eta_{tot\ j}}
\frac{\Delta_{(i-1)} x}{4})) \frac{\Gamma}{\Phi_*}
= 1
\ \ .
\end{equation}

When the last equation is solved then equation (\ref{sf})
expresses $N_i(\Delta x / 4)$
through $\Delta x$
and solves the problem.
Equation (\ref{ggg})
gives the expression for $g_i$
and  for the supersaturation as the function of time.

 One can estimate the relative error by the pseudo homogeneous case which leads to
\begin{equation} \label{93}
\frac{\mid N_{i}(\infty) - N_{final\ i}(\infty) \mid}{N_{i}(\infty)}
\leq \frac{\mid A -B \mid}{A} \sim 0.02
\end{equation}
The investigation of the process of nucleation
on  several sorts of heterogeneous centers
 is completed.

\section{Solution for heterogeneous centers with  quasi continuous activity}

\subsection{Iteration procedure}

The smoothness  of  $\eta(w)$ is supposed here. It
allows  to hope that the
cross influence of  exhaustion of one type of heterogeneous centers on
the cut-off of the droplets formation  on another sort of centers will
be not so strong and one can apply some modifications of the standard iteration
procedure.
 It can be constructed in a  following way: for  initial approximations
one can  choose
\begin{equation}
g_{0}(z,w) = 0 \ \ , \ \ \ \ \theta_{0} = 1 \ \ ;
\end{equation}
the recurrent procedure is defined according to
\begin{equation}
g_{i+1}(z,w) = G_w (g^{tot}_{i}, \theta_{i}(w) )
\ \ ,
\ \ \ \
g^{tot}_{i}(z) = \int dw g_{i}(z,w)
\ \ .
\end{equation}
\begin{equation} \label{****}
\theta_{i+1}(z,w) = S_w(g^{tot}_{i})
\end{equation}
The  monotonic properties of $G_w$, $\int dw$ and $S_w$
lead to the chains of inequalities analogous to
(\ref{chain})-(\ref{chain}), which
ensure
 estimates for precision of  approximations.
 Certainly, to calculate  iterations
 one has to use some expression for $\eta_{tot}(w)$. So, at first
one has to consider the limit cases and obtain some estimates.

One can see  the absence of $\eta_{tot}$ in the r.h.s. of
(\ref{****}). So, for  arbitrary $\eta_{tot}$ the power of exhaustion
will be determined  only by $w$. The centers with the high activity $w$
are almost exhausted during the process of the condensation. The centers with
the relatively small activity $w$ remain free. The intermediate region
has
rather
small size. To give the qualitative estimates one can  see that the
supersaturation
$\zeta$ appears only in the function $f_{\zeta}$. For
$f_{\zeta}$
one can obtain the following
estimate
\begin{equation} \label{*888}
f_{\zeta} \sim f_* \exp(-const x^{\epsilon}) \ \ \ \ \  3\le \epsilon \le 4
\end{equation}
This estimate  goes from the  obvious fact that the spectrum of the
droplets
must be  wider than the mono-disperse one and the intensity of nucleation
must
decrease in time.
On the base of
  the last estimates one can  see that the width $\Delta x$ of the spectrum
 is a well defined value.

The  rough approximation for  exhaustion of  heterogeneous
centers  can be obtained
from the first approximation in the  iteration procedure
\begin{equation}
\theta = \exp [ -  f_{*} \exp(\lambda w) n_{\infty} z ]
\end{equation}
and for the value at the end of nucleation period
\begin{equation}
\theta_{final} = \exp [ - f_{*} \exp(\lambda w) n_{\infty} \Delta x ]
\ \ .
\end{equation}
One can define $w_{<}$ and $w_{>}$ according to
\begin{equation}
w_{<} = w_{0} - \frac{\epsilon}{\lambda} \ \ ,
\ \ \ \  \
w_{>} = w_{0} + \frac{\epsilon}{\lambda} \ \ ,
\end{equation}
where
\begin{equation}
w_{0} = \frac{1}{\lambda} \ln (\frac{ \ln 2 }{ f_{*} n_{\infty} \Delta x } )
\ \ , \ \ \epsilon \sim 1 \ \ .
\end{equation}
These constructions are illustrated in Fig. 8.
\begin{figure}



{\small \it
\begin{center}
Figure 8.
\\
Function $\theta$ and boundaries $w_<$, $w_>$.
   \end{center}
}

\end{figure}
For $w > w_{>}$ almost all heterogeneous centers are exhausted. For $w<w_{<}$
almost all heterogeneous centers remain  free.
This remark  allows to rewrite
 the expression for $g^{tot}$ in the following form
\begin{equation}
g^{tot} = \int_{w_{<}}^{w_{>}} dw g(z,w)
+ \int_{w_>}^{\infty} \eta_{tot}(w) dw \frac{z^3}{n_{\infty}}
\ \ .
\end{equation}
The size of the intermediate region corresponds to a variation  of $w$
of
an order  $1/\lambda$ or
to the variation of $\Delta F$ in  one thermal unit.  Since  $\Delta F \gg 1$,
the relative variation of $\Delta F$ in the intermediate
region
is
 small. So, it is reasonable to put in this region
\begin{equation}
\eta_{tot}(w) = \eta_{*} = const \ \ .
\end{equation}
One can spread this approximation over the region $w>w_{>}$ because the
behavior of $\eta_{tot}$ for
 these centers isn't important.
 Only $\int \eta_{tot} dw $ over this region is essential.
 Certainly, one has to get the
boundary $w_{max}$ of this region by
\begin{equation}
\int_{w_{>}}^{w_{max}} \eta_{*} dw = \eta_{*}(w_{max}-w_{>}) =
\int_{w_{>}}^{\infty} \eta_{tot}(w) dw
\ \ .
\end{equation}
For $w<w_{<}$ the accuracy of this approximation is
not so essential because these centers remain
free.

After the scale transformations $\Gamma g^{tot}/\Phi_{*} \rightarrow G ; \ \ \
\Gamma g /\Phi_{*} \rightarrow g$
\begin{equation} \label{*a1}
g=\frac{\Gamma f_{*} \eta_{*}} {\Phi_{*}} \exp(\lambda w) \int_{0}^{z}
(z-x)^3 \exp(-G) \theta dx
\ \ ,
\end{equation}
\begin{equation} \label{*a2}
G = \int_{-\infty}^{w_{max}} dw g
\ \ ,
\end{equation}
\begin{equation} \label{*a3}
\theta = \exp [ - f_{*} n_{\infty} \exp(\lambda w) \int_{0}^{z} \exp(-G) dx ]
\ \ .
\end{equation}
By the  appropriate shift of $w$ one can put $w_{max}$ to the zero. In the
appropriate
 scale of $w$ one can put $\lambda = 1$.  In the appropriate scale of $z$
one puts the coefficient in (\ref{*a1})
 to unity.  As the result, only one
parameter - the coefficient
$
f_{*}^{3/4} \exp(3\lambda w_{max} / 4) \Phi_*^{1/4}/ (\Gamma^{1/4} \eta_*^{1/4})
$
remains
in the last equation. One can mark it by $A$
and get
\begin{equation} \label{*a4}
g=\exp(w) \int_{0}^{z} (z-x)^3 \exp(-G) \theta dx
\ \ ,
\end{equation}
\begin{equation} \label{*a5}
G = \int_{-\infty}^{0} dw g
\ \ ,
\end{equation}
\begin{equation} \label{*a6}
\theta = \exp[-A \exp(w) \int_{0}^{z} \exp(-G) dx ]
\ \ .
\end{equation}
An iteration procedure can be constructed as
\begin{equation}
g_{i+1} = \exp(w) \int_0^z (z-x)^3 \exp(-G_i) \theta_i dx                       \ \ ,
\end{equation}
\begin{equation}
G_i = \int_{-\infty}^0 dw g_i                                            \ \ ,
\end{equation}
\begin{equation}
\theta_{i+1} = \exp(-A \exp(w) \int_0^z \exp(-G_i) dx)                            \ \ ,
\end{equation}
\begin{equation}
g_0 = 0 \ \ , \ \  \  \theta_0 = 1                                                        \ \ .
\end{equation}
In the first iteration
\begin{equation}
g_1 = \exp(w) \frac{z^4}{4} \ \ ,
\end{equation}
\begin{equation}
G_1 = \int_{-\infty}^0 dw g_1 = \frac{z^4}{4} \ \ ,
\end{equation}
\begin{equation}
\theta_1 = \exp( - A \exp(w) z) \ \ .
\end{equation}
The second approximation gives for $\theta$ the following result
\begin{equation}
\theta_{2} = \exp[-A \exp(w) \int_{0}^{z} \exp(-\frac{z^4}{4}) dz]
\ \ ,
\end{equation}
then for  the final value
\begin{equation}
\theta_{final\ 2}= \exp[-A \exp(w)  1.28^{1/4}] \ \
\end{equation}
and for  $N^{tot}$
\begin{equation}
N^{tot} = \int_{-\infty}^{0}[1-\exp(-Aexp(w) 1.28^{1/4})] dw    \eta_*
\ \ .
\end{equation}
For $g$ in the second approximation
\begin{equation}
g_{2} = \exp(w) \int_{0}^{z} (z-x)^3 \exp(-A \exp(w) x ) \exp(- x^4/4) dx
\ \ .
\end{equation}
The analytical expression for $g_2$ and  further iterations can not be
calculated.

\subsection{Universal solution}

The system of equations (\ref{*a4}) - (\ref{*a6}) doesn't allow the universal
solution as in the case of homogeneous condensation when  (\ref{*a4})
hasn't $\theta$ in the r.s.h.. But the operator in the r.h.s. of (\ref{*a4})
ensures
 a rapid convergence of the iterations. The worst situation for
the iteration convergence is  $A=0$. In this situation one has
after rescaling the universal
system
\begin{equation}
g = \exp(w) \int_{0}^{z} (z-x)^3 \exp(-G) dx
\ \ ,
\end{equation}
\begin{equation}
G=\int_{-\infty}^{0} g dw
\ \ .
\end{equation}
This system has the universal solution\footnote{The trivial integration over
$w$ leads to the equation analogous to the homogeneous case.} which will be marked by $G_{0}$.
This universal form of the size spectrum is shown in  Fig. 9.
\begin{figure}

\begin{picture}(450,250)
\put(60,10){\special{em: moveto}}
\put(290,10){\special{em: lineto}}
\put(290,240){\special{em: lineto}}
\put(60,240){\special{em: lineto}}
\put(60,10){\special{em: lineto}}
\put(65,15){\special{em: moveto}}
\put(285,15){\special{em: lineto}}
\put(285,235){\special{em: lineto}}
\put(65,235){\special{em: lineto}}
\put(65,15){\special{em: lineto}}
\put(100,50){\vector(1,0){150}}
\put(100,50){\vector(0,1){150}}
\put(150,50){\vector(0,1){3}}
\put(200,50){\vector(0,1){3}}
\put(100,100){\vector(1,0){3}}
\put(100,150){\vector(1,0){3}}
\put(90,40){$0$}
\put(148,40){$0.5$}
\put(198,40){$1$}
\put(82,98){$0.5$}
\put(90,148){$1$}
\put(108,208){$f \ \ (in \ relative \ units ) $}
\put(255,55){$x$}
\put(150,180){HOMOGENEOUS}
\put(152,160){UNIVERSAL}
\put(156,140){SPECTRUM}
\put(100,150){\special{em: moveto}}
\put(100.50,150.00){\special{em: lineto}}
\put(101.00,150.00){\special{em: lineto}}
\put(101.50,150.00){\special{em: lineto}}
\put(102.00,150.00){\special{em: lineto}}
\put(102.50,150.00){\special{em: lineto}}
\put(103.00,150.00){\special{em: lineto}}
\put(103.50,150.00){\special{em: lineto}}
\put(104.00,150.00){\special{em: lineto}}
\put(104.50,150.00){\special{em: lineto}}
\put(105.00,150.00){\special{em: lineto}}
\put(105.50,150.00){\special{em: lineto}}
\put(106.00,150.00){\special{em: lineto}}
\put(106.50,150.00){\special{em: lineto}}
\put(107.00,149.99){\special{em: lineto}}
\put(107.50,149.99){\special{em: lineto}}
\put(108.00,149.99){\special{em: lineto}}
\put(108.50,149.99){\special{em: lineto}}
\put(109.00,149.98){\special{em: lineto}}
\put(109.50,149.98){\special{em: lineto}}
\put(110.00,149.97){\special{em: lineto}}
\put(110.50,149.96){\special{em: lineto}}
\put(111.00,149.96){\special{em: lineto}}
\put(111.50,149.95){\special{em: lineto}}
\put(112.00,149.94){\special{em: lineto}}
\put(112.50,149.92){\special{em: lineto}}
\put(113.00,149.91){\special{em: lineto}}
\put(113.50,149.89){\special{em: lineto}}
\put(114.00,149.88){\special{em: lineto}}
\put(114.50,149.86){\special{em: lineto}}
\put(115.00,149.84){\special{em: lineto}}
\put(115.50,149.81){\special{em: lineto}}
\put(116.00,149.78){\special{em: lineto}}
\put(116.50,149.75){\special{em: lineto}}
\put(117.00,149.72){\special{em: lineto}}
\put(117.50,149.69){\special{em: lineto}}
\put(118.00,149.65){\special{em: lineto}}
\put(118.50,149.60){\special{em: lineto}}
\put(119.00,149.56){\special{em: lineto}}
\put(119.50,149.51){\special{em: lineto}}
\put(120.00,149.45){\special{em: lineto}}
\put(120.50,149.39){\special{em: lineto}}
\put(121.00,149.33){\special{em: lineto}}
\put(121.50,149.26){\special{em: lineto}}
\put(122.00,149.19){\special{em: lineto}}
\put(122.50,149.11){\special{em: lineto}}
\put(123.00,149.02){\special{em: lineto}}
\put(123.50,148.93){\special{em: lineto}}
\put(124.00,148.84){\special{em: lineto}}
\put(124.50,148.74){\special{em: lineto}}
\put(125.00,148.63){\special{em: lineto}}
\put(125.50,148.51){\special{em: lineto}}
\put(126.00,148.39){\special{em: lineto}}
\put(126.50,148.26){\special{em: lineto}}
\put(127.00,148.12){\special{em: lineto}}
\put(127.50,147.97){\special{em: lineto}}
\put(128.00,147.82){\special{em: lineto}}
\put(128.50,147.66){\special{em: lineto}}
\put(129.00,147.49){\special{em: lineto}}
\put(129.50,147.31){\special{em: lineto}}
\put(130.00,147.12){\special{em: lineto}}
\put(130.50,146.92){\special{em: lineto}}
\put(131.00,146.71){\special{em: lineto}}
\put(131.50,146.49){\special{em: lineto}}
\put(132.00,146.26){\special{em: lineto}}
\put(132.50,146.02){\special{em: lineto}}
\put(133.00,145.77){\special{em: lineto}}
\put(133.50,145.51){\special{em: lineto}}
\put(134.00,145.23){\special{em: lineto}}
\put(134.50,144.95){\special{em: lineto}}
\put(135.00,144.65){\special{em: lineto}}
\put(135.50,144.34){\special{em: lineto}}
\put(136.00,144.02){\special{em: lineto}}
\put(136.50,143.68){\special{em: lineto}}
\put(137.00,143.33){\special{em: lineto}}
\put(137.50,142.97){\special{em: lineto}}
\put(138.00,142.60){\special{em: lineto}}
\put(138.50,142.21){\special{em: lineto}}
\put(139.00,141.80){\special{em: lineto}}
\put(139.50,141.39){\special{em: lineto}}
\put(140.00,140.95){\special{em: lineto}}
\put(140.50,140.51){\special{em: lineto}}
\put(141.00,140.05){\special{em: lineto}}
\put(141.50,139.57){\special{em: lineto}}
\put(142.00,139.08){\special{em: lineto}}
\put(142.50,138.57){\special{em: lineto}}
\put(143.00,138.05){\special{em: lineto}}
\put(143.50,137.51){\special{em: lineto}}
\put(144.00,136.96){\special{em: lineto}}
\put(144.50,136.39){\special{em: lineto}}
\put(145.00,135.81){\special{em: lineto}}
\put(145.50,135.21){\special{em: lineto}}
\put(146.00,134.59){\special{em: lineto}}
\put(146.50,133.96){\special{em: lineto}}
\put(147.00,133.31){\special{em: lineto}}
\put(147.50,132.65){\special{em: lineto}}
\put(148.00,131.97){\special{em: lineto}}
\put(148.50,131.27){\special{em: lineto}}
\put(149.00,130.56){\special{em: lineto}}
\put(149.50,129.83){\special{em: lineto}}
\put(150.00,129.09){\special{em: lineto}}
\put(150.50,128.33){\special{em: lineto}}
\put(151.00,127.56){\special{em: lineto}}
\put(151.50,126.77){\special{em: lineto}}
\put(152.00,125.96){\special{em: lineto}}
\put(152.50,125.14){\special{em: lineto}}
\put(153.00,124.31){\special{em: lineto}}
\put(153.50,123.46){\special{em: lineto}}
\put(154.00,122.60){\special{em: lineto}}
\put(154.50,121.72){\special{em: lineto}}
\put(155.00,120.83){\special{em: lineto}}
\put(155.50,119.93){\special{em: lineto}}
\put(156.00,119.01){\special{em: lineto}}
\put(156.50,118.08){\special{em: lineto}}
\put(157.00,117.14){\special{em: lineto}}
\put(157.50,116.19){\special{em: lineto}}
\put(158.00,115.23){\special{em: lineto}}
\put(158.50,114.25){\special{em: lineto}}
\put(159.00,113.27){\special{em: lineto}}
\put(159.50,112.28){\special{em: lineto}}
\put(160.00,111.27){\special{em: lineto}}
\put(160.50,110.26){\special{em: lineto}}
\put(161.00,109.24){\special{em: lineto}}
\put(161.50,108.22){\special{em: lineto}}
\put(162.00,107.19){\special{em: lineto}}
\put(162.50,106.15){\special{em: lineto}}
\put(163.00,105.10){\special{em: lineto}}
\put(163.50,104.05){\special{em: lineto}}
\put(164.00,103.00){\special{em: lineto}}
\put(164.50,101.94){\special{em: lineto}}
\put(165.00,100.88){\special{em: lineto}}
\put(165.50,99.82){\special{em: lineto}}
\put(166.00,98.76){\special{em: lineto}}
\put(166.50,97.69){\special{em: lineto}}
\put(167.00,96.63){\special{em: lineto}}
\put(167.50,95.57){\special{em: lineto}}
\put(168.00,94.50){\special{em: lineto}}
\put(168.50,93.44){\special{em: lineto}}
\put(169.00,92.39){\special{em: lineto}}
\put(169.50,91.34){\special{em: lineto}}
\put(170.00,90.29){\special{em: lineto}}
\put(170.50,89.24){\special{em: lineto}}
\put(171.00,88.21){\special{em: lineto}}
\put(171.50,87.18){\special{em: lineto}}
\put(172.00,86.15){\special{em: lineto}}
\put(172.50,85.14){\special{em: lineto}}
\put(173.00,84.13){\special{em: lineto}}
\put(173.50,83.13){\special{em: lineto}}
\put(174.00,82.14){\special{em: lineto}}
\put(174.50,81.17){\special{em: lineto}}
\put(175.00,80.20){\special{em: lineto}}
\put(175.50,79.25){\special{em: lineto}}
\put(176.00,78.31){\special{em: lineto}}
\put(176.50,77.38){\special{em: lineto}}
\put(177.00,76.46){\special{em: lineto}}
\put(177.50,75.56){\special{em: lineto}}
\put(178.00,74.68){\special{em: lineto}}
\put(178.50,73.80){\special{em: lineto}}
\put(179.00,72.95){\special{em: lineto}}
\put(179.50,72.11){\special{em: lineto}}
\put(180.00,71.28){\special{em: lineto}}
\put(180.50,70.48){\special{em: lineto}}
\put(181.00,69.68){\special{em: lineto}}
\put(181.50,68.91){\special{em: lineto}}
\put(182.00,68.15){\special{em: lineto}}
\put(182.50,67.42){\special{em: lineto}}
\put(183.00,66.69){\special{em: lineto}}
\put(183.50,65.99){\special{em: lineto}}
\put(184.00,65.31){\special{em: lineto}}
\put(184.50,64.64){\special{em: lineto}}
\put(185.00,63.99){\special{em: lineto}}
\put(185.50,63.36){\special{em: lineto}}
\put(186.00,62.75){\special{em: lineto}}
\put(186.50,62.16){\special{em: lineto}}
\put(187.00,61.59){\special{em: lineto}}
\put(187.50,61.03){\special{em: lineto}}
\put(188.00,60.49){\special{em: lineto}}
\put(188.50,59.97){\special{em: lineto}}
\put(189.00,59.47){\special{em: lineto}}
\put(189.50,58.99){\special{em: lineto}}
\put(190.00,58.52){\special{em: lineto}}
\put(190.50,58.07){\special{em: lineto}}
\put(191.00,57.64){\special{em: lineto}}
\put(191.50,57.22){\special{em: lineto}}
\put(192.00,56.83){\special{em: lineto}}
\put(192.50,56.44){\special{em: lineto}}
\put(193.00,56.08){\special{em: lineto}}
\put(193.50,55.73){\special{em: lineto}}
\put(194.00,55.39){\special{em: lineto}}
\put(194.50,55.07){\special{em: lineto}}
\put(195.00,54.77){\special{em: lineto}}
\put(195.50,54.48){\special{em: lineto}}
\put(196.00,54.20){\special{em: lineto}}
\put(196.50,53.93){\special{em: lineto}}
\put(197.00,53.68){\special{em: lineto}}
\put(197.50,53.45){\special{em: lineto}}
\put(198.00,53.22){\special{em: lineto}}
\put(198.50,53.01){\special{em: lineto}}
\put(199.00,52.80){\special{em: lineto}}
\put(199.50,52.61){\special{em: lineto}}
\put(200.00,52.43){\special{em: lineto}}
\put(200.50,52.26){\special{em: lineto}}
\put(201.00,52.10){\special{em: lineto}}
\put(201.50,51.95){\special{em: lineto}}
\put(202.00,51.81){\special{em: lineto}}
\put(202.50,51.67){\special{em: lineto}}
\put(203.00,51.55){\special{em: lineto}}
\put(203.50,51.43){\special{em: lineto}}
\put(204.00,51.32){\special{em: lineto}}
\put(204.50,51.22){\special{em: lineto}}
\put(205.00,51.12){\special{em: lineto}}
\put(205.50,51.03){\special{em: lineto}}
\put(206.00,50.95){\special{em: lineto}}
\put(206.50,50.87){\special{em: lineto}}
\put(207.00,50.80){\special{em: lineto}}
\put(207.50,50.73){\special{em: lineto}}
\put(208.00,50.67){\special{em: lineto}}
\put(208.50,50.61){\special{em: lineto}}
\put(209.00,50.56){\special{em: lineto}}
\put(209.50,50.51){\special{em: lineto}}
\put(210.00,50.47){\special{em: lineto}}
\put(210.50,50.42){\special{em: lineto}}
\put(211.00,50.38){\special{em: lineto}}
\put(211.50,50.35){\special{em: lineto}}
\put(212.00,50.32){\special{em: lineto}}
\put(212.50,50.29){\special{em: lineto}}
\put(213.00,50.26){\special{em: lineto}}
\put(213.50,50.23){\special{em: lineto}}
\put(214.00,50.21){\special{em: lineto}}
\put(214.50,50.19){\special{em: lineto}}
\put(215.00,50.17){\special{em: lineto}}
\put(215.50,50.15){\special{em: lineto}}
\put(216.00,50.14){\special{em: lineto}}
\put(216.50,50.12){\special{em: lineto}}
\put(217.00,50.11){\special{em: lineto}}
\put(217.50,50.10){\special{em: lineto}}
\put(218.00,50.09){\special{em: lineto}}
\put(218.50,50.08){\special{em: lineto}}
\put(219.00,50.07){\special{em: lineto}}
\put(219.50,50.06){\special{em: lineto}}
\put(220.00,50.06){\special{em: lineto}}
\put(220.50,50.05){\special{em: lineto}}
\put(221.00,50.04){\special{em: lineto}}
\put(221.50,50.04){\special{em: lineto}}
\put(222.00,50.03){\special{em: lineto}}
\put(222.50,50.03){\special{em: lineto}}
\put(223.00,50.03){\special{em: lineto}}
\put(223.50,50.02){\special{em: lineto}}
\put(224.00,50.02){\special{em: lineto}}
\put(224.50,50.02){\special{em: lineto}}
\put(225.00,50.02){\special{em: lineto}}
\put(225.50,50.01){\special{em: lineto}}
\put(226.00,50.01){\special{em: lineto}}
\put(226.50,50.01){\special{em: lineto}}
\put(227.00,50.01){\special{em: lineto}}
\put(227.50,50.01){\special{em: lineto}}
\put(228.00,50.01){\special{em: lineto}}
\put(228.50,50.01){\special{em: lineto}}
\put(229.00,50.01){\special{em: lineto}}
\put(229.50,50.00){\special{em: lineto}}
\put(230.00,50.00){\special{em: lineto}}
\put(230.50,50.00){\special{em: lineto}}
\put(231.00,50.00){\special{em: lineto}}
\put(231.50,50.00){\special{em: lineto}}
\put(232.00,50.00){\special{em: lineto}}
\put(232.50,50.00){\special{em: lineto}}
\put(233.00,50.00){\special{em: lineto}}
\put(233.50,50.00){\special{em: lineto}}
\put(234.00,50.00){\special{em: lineto}}
\put(234.50,50.00){\special{em: lineto}}
\put(235.00,50.00){\special{em: lineto}}
\put(235.50,50.00){\special{em: lineto}}
\put(236.00,50.00){\special{em: lineto}}
\put(236.50,50.00){\special{em: lineto}}
\put(237.00,50.00){\special{em: lineto}}
\put(237.50,50.00){\special{em: lineto}}
\put(238.00,50.00){\special{em: lineto}}
\put(238.50,50.00){\special{em: lineto}}
\put(239.00,50.00){\special{em: lineto}}
\put(239.50,50.00){\special{em: lineto}}
\put(240.00,50.00){\special{em: lineto}}
\put(240.50,50.00){\special{em: lineto}}
\put(241.00,50.00){\special{em: lineto}}
\put(241.50,50.00){\special{em: lineto}}
\put(242.00,50.00){\special{em: lineto}}
\put(242.50,50.00){\special{em: lineto}}
\put(243.00,50.00){\special{em: lineto}}
\put(243.50,50.00){\special{em: lineto}}
\put(244.00,50.00){\special{em: lineto}}
\put(244.50,50.00){\special{em: lineto}}
\put(245.00,50.00){\special{em: lineto}}
\put(245.50,50.00){\special{em: lineto}}
\put(246.00,50.00){\special{em: lineto}}
\put(246.50,50.00){\special{em: lineto}}
\put(247.00,50.00){\special{em: lineto}}
\put(247.50,50.00){\special{em: lineto}}
\put(248.00,50.00){\special{em: lineto}}
\put(248.50,50.00){\special{em: lineto}}
\put(249.00,50.00){\special{em: lineto}}
\put(249.50,50.00){\special{em: lineto}}
\put(250.00,50.00){\special{em: lineto}}
\put(250.50,50.00){\special{em: lineto}}
\end{picture}

{\small \it
\begin{center}
Figure 9.
\\
Universal form of spectrum in the pseudo homogeneous case.
   \end{center}
}

\end{figure}

The  evolution is determined  by the first three momentums (and
the zero one) of the
distribution
function
\begin{equation} \label{*cx0}
\mu_{i}(z) = \int_{0}^{z} x^{i} \exp(-G) \theta dx \ \ ,
\ \ \ \
i=0,1,2,3
\end{equation}
and in the pseudo homogeneous case it is determined by
\begin{equation} \label{*cx0dd}
\mu_{i}(z) = \int_{0}^{z} x^{i} \exp(-G_0) dx \ \ .
\end{equation}

After the end of the  short period of  intensive formation of droplets one can
substitute
in (\ref{*cx0}) and (\ref{*cx0dd}) $\infty$ instead of $z$ in the region of
the integration. So,  the further evolution will be determined  by
four constants\footnote{
The main influence is due to $\mu_{0}$.}
$\mu_{i}(\infty) \ i=0,1,2,3 $. If one takes for $G$   the
universal
solution $G_{0}$
(when $A=0$) then the values of $\mu_{i}$ are the universal constants which can
be obtained by the unique numerical solution of the last system.

Now one has to return to the  iteration procedure.
One can use $G_{0}$ as the initial approximation in the iteration procedure and get
\begin{equation} \label{*d1}
\theta_{1} = \exp[-A \exp(w) \int_{0}^{z} \exp(-G_{0}) dx]
\ \ ,
\end{equation}
\begin{equation}
g_{2} = \exp(w) \int_{0}^{z} (z-x)^3 \exp(-G_{0}) \exp(-A \exp(w) \int_{0}^{x}
\exp(-G_{0}) dx') dx
\ \ ,
\end{equation}
\begin{equation}
G_{2} = \int_{-\infty}^{0}
\exp(w) \int_{0}^{z} (z-x)^3 \exp(-G_{0}) \exp(-A \exp(w) \int_{0}^{x}
\exp(-G_{0}) dx') dx
dw
\ \ .
\end{equation}
The decomposition of exponents gives
\begin{equation}
G_2 = \sum_{i=0}^{\infty} \frac{(-A)^i}{(i+1)!} P_i(z)
\ \ ,
\end{equation}
where
\begin{equation}
P_{i} = \int_{0}^{z} (z-x)^3 \exp(-G_{0}) J_{00}^{i}(x) dx
\ \ ,
\end{equation}
\begin{equation}
J_{00}(x) = \int_{0}^{x} \exp(-G_{0})dx
\ \ .
\end{equation}
Then  the function $B(z) = \int_0^z \exp (-G_2(x)) dx$
at $z=\infty$ will be the following
\begin{equation}
B(z=\infty) = A \int_{0}^{\infty} \exp(-G_2) dx =
A \int_0^{\infty}
\exp(-P_0(z))
\exp(-\sum_{i=1}^{\infty} \frac{(-A)^i}{(i+1)!} P_i(z)) dz
\ \ .
\end{equation}
The decomposition of the last  exponent leads to
\begin{equation}
B(z=\infty) = \prod_{i=1}^{\infty} \sum_{j=0}^{\infty} \frac{(-1)^j}{j!} A
\frac{(-A)^{ij}}{((i+1)!)^j} C_{ij} \ \ ,
\end{equation}
where
\begin{equation}
C_{ij} =
\int_{0}^{\infty} \exp(-P_{0}(x)) P_{i}^{j}  dx
\end{equation}
are the universal constants.
For the total number of the droplets with a given activity
\begin{equation} \label{*d9}
N_{tot}(w) = \eta_{*} (1 - \exp(-B(z=\infty) \exp(w))) \ \ .
\end{equation}

In the same manner one can take into account the deviation of $\eta_{*}$
from a constant value.
After the decomposition of $\eta_{*}$ into  Tailor series one can  take
the
initial approximation  as
$G_{0}$ and act in the manner similar to (\ref{*d1}) -
(\ref{*d9}). As a result, some expansions in the powers of the parameter
$A$,
the  derivatives  of $\eta_{tot}$ and the universal constants can be seen.

Fig. 10 illustrates these constructions.
\begin{figure}

\begin{picture}(450,250)
\put(60,10){\special{em: moveto}}
\put(290,10){\special{em: lineto}}
\put(290,240){\special{em: lineto}}
\put(60,240){\special{em: lineto}}
\put(60,10){\special{em: lineto}}
\put(65,15){\special{em: moveto}}
\put(285,15){\special{em: lineto}}
\put(285,235){\special{em: lineto}}
\put(65,235){\special{em: lineto}}
\put(65,15){\special{em: lineto}}
\put(100,50){\vector(1,0){150}}
\put(100,50){\vector(0,1){150}}
\put(150,50){\vector(0,1){3}}
\put(200,50){\vector(0,1){3}}
\put(100,100){\vector(1,0){3}}
\put(100,150){\vector(1,0){3}}
\put(100,30){$0$}
\put(148,40){$5$}
\put(198,40){$10$}
\put(80,48){$0$}
\put(82,98){$0.1$}
\put(82,148){$0.2$}
\put(108,208){$I_{00}-1.3$}
\put(255,55){$A$}
\put(101.00,49.08){\special{em: moveto}}
\put(102.00,51.14){\special{em: lineto}}
\put(103.00,53.17){\special{em: lineto}}
\put(104.00,55.18){\special{em: lineto}}
\put(105.00,57.18){\special{em: lineto}}
\put(106.00,59.15){\special{em: lineto}}
\put(107.00,61.10){\special{em: lineto}}
\put(108.00,63.03){\special{em: lineto}}
\put(109.00,64.94){\special{em: lineto}}
\put(110.00,66.83){\special{em: lineto}}
\put(111.00,68.71){\special{em: lineto}}
\put(112.00,70.56){\special{em: lineto}}
\put(113.00,72.39){\special{em: lineto}}
\put(114.00,74.20){\special{em: lineto}}
\put(115.00,75.99){\special{em: lineto}}
\put(116.00,77.76){\special{em: lineto}}
\put(117.00,79.52){\special{em: lineto}}
\put(118.00,81.25){\special{em: lineto}}
\put(119.00,82.96){\special{em: lineto}}
\put(120.00,84.66){\special{em: lineto}}
\put(121.00,86.33){\special{em: lineto}}
\put(122.00,87.99){\special{em: lineto}}
\put(123.00,89.63){\special{em: lineto}}
\put(124.00,91.25){\special{em: lineto}}
\put(125.00,92.86){\special{em: lineto}}
\put(126.00,94.44){\special{em: lineto}}
\put(127.00,96.01){\special{em: lineto}}
\put(128.00,97.56){\special{em: lineto}}
\put(129.00,99.10){\special{em: lineto}}
\put(130.00,100.61){\special{em: lineto}}
\put(131.00,102.11){\special{em: lineto}}
\put(132.00,103.60){\special{em: lineto}}
\put(133.00,105.07){\special{em: lineto}}
\put(134.00,106.52){\special{em: lineto}}
\put(135.00,107.95){\special{em: lineto}}
\put(136.00,109.37){\special{em: lineto}}
\put(137.00,110.78){\special{em: lineto}}
\put(138.00,112.17){\special{em: lineto}}
\put(139.00,113.54){\special{em: lineto}}
\put(140.00,114.90){\special{em: lineto}}
\put(141.00,116.25){\special{em: lineto}}
\put(142.00,117.58){\special{em: lineto}}
\put(143.00,118.90){\special{em: lineto}}
\put(144.00,120.20){\special{em: lineto}}
\put(145.00,121.49){\special{em: lineto}}
\put(146.00,122.77){\special{em: lineto}}
\put(147.00,124.04){\special{em: lineto}}
\put(148.00,125.29){\special{em: lineto}}
\put(149.00,126.52){\special{em: lineto}}
\put(150.00,127.75){\special{em: lineto}}
\put(151.00,128.96){\special{em: lineto}}
\put(152.00,130.16){\special{em: lineto}}
\put(153.00,131.35){\special{em: lineto}}
\put(154.00,132.53){\special{em: lineto}}
\put(155.00,133.69){\special{em: lineto}}
\put(156.00,134.84){\special{em: lineto}}
\put(157.00,135.98){\special{em: lineto}}
\put(158.00,137.11){\special{em: lineto}}
\put(159.00,138.23){\special{em: lineto}}
\put(160.00,139.34){\special{em: lineto}}
\put(161.00,140.44){\special{em: lineto}}
\put(162.00,141.53){\special{em: lineto}}
\put(163.00,142.60){\special{em: lineto}}
\put(164.00,143.67){\special{em: lineto}}
\put(165.00,144.72){\special{em: lineto}}
\put(166.00,145.77){\special{em: lineto}}
\put(167.00,146.81){\special{em: lineto}}
\put(168.00,147.83){\special{em: lineto}}
\put(169.00,148.85){\special{em: lineto}}
\put(170.00,149.86){\special{em: lineto}}
\put(171.00,150.85){\special{em: lineto}}
\put(172.00,151.84){\special{em: lineto}}
\put(173.00,152.82){\special{em: lineto}}
\put(174.00,153.79){\special{em: lineto}}
\put(175.00,154.75){\special{em: lineto}}
\put(176.00,155.71){\special{em: lineto}}
\put(177.00,156.65){\special{em: lineto}}
\put(178.00,157.59){\special{em: lineto}}
\put(179.00,158.52){\special{em: lineto}}
\put(180.00,159.44){\special{em: lineto}}
\put(181.00,160.35){\special{em: lineto}}
\put(182.00,161.25){\special{em: lineto}}
\put(183.00,162.15){\special{em: lineto}}
\put(184.00,163.04){\special{em: lineto}}
\put(185.00,163.92){\special{em: lineto}}
\put(186.00,164.79){\special{em: lineto}}
\put(187.00,165.66){\special{em: lineto}}
\put(188.00,166.51){\special{em: lineto}}
\put(189.00,167.36){\special{em: lineto}}
\put(190.00,168.21){\special{em: lineto}}
\put(191.00,169.04){\special{em: lineto}}
\put(192.00,169.87){\special{em: lineto}}
\put(193.00,170.70){\special{em: lineto}}
\put(194.00,171.51){\special{em: lineto}}
\put(195.00,172.32){\special{em: lineto}}
\put(196.00,173.13){\special{em: lineto}}
\put(197.00,173.92){\special{em: lineto}}
\put(198.00,174.71){\special{em: lineto}}
\put(199.00,175.49){\special{em: lineto}}
\put(200.00,176.27){\special{em: lineto}}
\put(201.00,177.04){\special{em: lineto}}
\put(202.00,177.81){\special{em: lineto}}
\put(203.00,178.57){\special{em: lineto}}
\put(204.00,179.32){\special{em: lineto}}
\put(205.00,180.07){\special{em: lineto}}
\put(206.00,180.81){\special{em: lineto}}
\put(207.00,181.54){\special{em: lineto}}
\put(208.00,182.27){\special{em: lineto}}
\put(209.00,183.00){\special{em: lineto}}
\put(210.00,183.71){\special{em: lineto}}
\put(211.00,184.43){\special{em: lineto}}
\put(212.00,185.13){\special{em: lineto}}
\put(213.00,185.84){\special{em: lineto}}
\put(214.00,186.53){\special{em: lineto}}
\put(215.00,187.23){\special{em: lineto}}
\put(216.00,187.91){\special{em: lineto}}
\put(217.00,188.59){\special{em: lineto}}
\put(218.00,189.27){\special{em: lineto}}
\put(219.00,189.94){\special{em: lineto}}
\put(220.00,190.61){\special{em: lineto}}
\put(221.00,191.27){\special{em: lineto}}
\put(222.00,191.93){\special{em: lineto}}
\put(223.00,192.58){\special{em: lineto}}
\put(224.00,193.23){\special{em: lineto}}
\put(225.00,193.87){\special{em: lineto}}
\put(226.00,194.51){\special{em: lineto}}
\put(227.00,195.14){\special{em: lineto}}
\put(228.00,195.77){\special{em: lineto}}
\put(229.00,196.40){\special{em: lineto}}
\put(230.00,197.02){\special{em: lineto}}
\put(231.00,197.63){\special{em: lineto}}
\put(232.00,198.25){\special{em: lineto}}
\put(233.00,198.85){\special{em: lineto}}
\put(234.00,199.46){\special{em: lineto}}
\put(235.00,200.06){\special{em: lineto}}
\put(236.00,200.65){\special{em: lineto}}
\put(237.00,201.24){\special{em: lineto}}
\put(238.00,201.83){\special{em: lineto}}
\put(239.00,202.41){\special{em: lineto}}
\put(240.00,202.99){\special{em: lineto}}
\put(241.00,203.57){\special{em: lineto}}
\put(242.00,204.14){\special{em: lineto}}
\put(243.00,204.71){\special{em: lineto}}
\put(244.00,205.27){\special{em: lineto}}
\put(245.00,205.83){\special{em: lineto}}
\put(246.00,206.39){\special{em: lineto}}
\put(247.00,206.94){\special{em: lineto}}
\put(248.00,207.49){\special{em: lineto}}
\put(249.00,208.04){\special{em: lineto}}
\put(250.00,208.58){\special{em: lineto}}
\put(101.00,44.13){\special{em: moveto}}
\put(102.00,46.03){\special{em: lineto}}
\put(103.00,47.93){\special{em: lineto}}
\put(104.00,49.83){\special{em: lineto}}
\put(105.00,51.73){\special{em: lineto}}
\put(106.00,53.62){\special{em: lineto}}
\put(107.00,55.50){\special{em: lineto}}
\put(108.00,57.38){\special{em: lineto}}
\put(109.00,59.25){\special{em: lineto}}
\put(110.00,61.10){\special{em: lineto}}
\put(111.00,62.94){\special{em: lineto}}
\put(112.00,64.77){\special{em: lineto}}
\put(113.00,66.59){\special{em: lineto}}
\put(114.00,68.39){\special{em: lineto}}
\put(115.00,70.18){\special{em: lineto}}
\put(116.00,71.95){\special{em: lineto}}
\put(117.00,73.71){\special{em: lineto}}
\put(118.00,75.45){\special{em: lineto}}
\put(119.00,77.18){\special{em: lineto}}
\put(120.00,78.88){\special{em: lineto}}
\put(121.00,80.58){\special{em: lineto}}
\put(122.00,82.25){\special{em: lineto}}
\put(123.00,83.91){\special{em: lineto}}
\put(124.00,85.55){\special{em: lineto}}
\put(125.00,87.17){\special{em: lineto}}
\put(126.00,88.78){\special{em: lineto}}
\put(127.00,90.37){\special{em: lineto}}
\put(128.00,91.94){\special{em: lineto}}
\put(129.00,93.50){\special{em: lineto}}
\put(130.00,95.04){\special{em: lineto}}
\put(131.00,96.56){\special{em: lineto}}
\put(132.00,98.07){\special{em: lineto}}
\put(133.00,99.56){\special{em: lineto}}
\put(134.00,101.04){\special{em: lineto}}
\put(135.00,102.50){\special{em: lineto}}
\put(136.00,103.94){\special{em: lineto}}
\put(137.00,105.37){\special{em: lineto}}
\put(138.00,106.79){\special{em: lineto}}
\put(139.00,108.19){\special{em: lineto}}
\put(140.00,109.57){\special{em: lineto}}
\put(141.00,110.94){\special{em: lineto}}
\put(142.00,112.30){\special{em: lineto}}
\put(143.00,113.64){\special{em: lineto}}
\put(144.00,114.96){\special{em: lineto}}
\put(145.00,116.28){\special{em: lineto}}
\put(146.00,117.57){\special{em: lineto}}
\put(147.00,118.86){\special{em: lineto}}
\put(148.00,120.13){\special{em: lineto}}
\put(149.00,121.39){\special{em: lineto}}
\put(150.00,122.64){\special{em: lineto}}
\put(151.00,123.87){\special{em: lineto}}
\put(152.00,125.09){\special{em: lineto}}
\put(153.00,126.30){\special{em: lineto}}
\put(154.00,127.50){\special{em: lineto}}
\put(155.00,128.68){\special{em: lineto}}
\put(156.00,129.86){\special{em: lineto}}
\put(157.00,131.02){\special{em: lineto}}
\put(158.00,132.17){\special{em: lineto}}
\put(159.00,133.31){\special{em: lineto}}
\put(160.00,134.44){\special{em: lineto}}
\put(161.00,135.55){\special{em: lineto}}
\put(162.00,136.66){\special{em: lineto}}
\put(163.00,137.75){\special{em: lineto}}
\put(164.00,138.84){\special{em: lineto}}
\put(165.00,139.91){\special{em: lineto}}
\put(166.00,140.97){\special{em: lineto}}
\put(167.00,142.03){\special{em: lineto}}
\put(168.00,143.07){\special{em: lineto}}
\put(169.00,144.11){\special{em: lineto}}
\put(170.00,145.13){\special{em: lineto}}
\put(171.00,146.15){\special{em: lineto}}
\put(172.00,147.15){\special{em: lineto}}
\put(173.00,148.15){\special{em: lineto}}
\put(174.00,149.14){\special{em: lineto}}
\put(175.00,150.12){\special{em: lineto}}
\put(176.00,151.08){\special{em: lineto}}
\put(177.00,152.05){\special{em: lineto}}
\put(178.00,153.00){\special{em: lineto}}
\put(179.00,153.94){\special{em: lineto}}
\put(180.00,154.88){\special{em: lineto}}
\put(181.00,155.81){\special{em: lineto}}
\put(182.00,156.72){\special{em: lineto}}
\put(183.00,157.64){\special{em: lineto}}
\put(184.00,158.54){\special{em: lineto}}
\put(185.00,159.43){\special{em: lineto}}
\put(186.00,160.32){\special{em: lineto}}
\put(187.00,161.20){\special{em: lineto}}
\put(188.00,162.08){\special{em: lineto}}
\put(189.00,162.94){\special{em: lineto}}
\put(190.00,163.80){\special{em: lineto}}
\put(191.00,164.65){\special{em: lineto}}
\put(192.00,165.49){\special{em: lineto}}
\put(193.00,166.33){\special{em: lineto}}
\put(194.00,167.16){\special{em: lineto}}
\put(195.00,167.98){\special{em: lineto}}
\put(196.00,168.80){\special{em: lineto}}
\put(197.00,169.61){\special{em: lineto}}
\put(198.00,170.41){\special{em: lineto}}
\put(199.00,171.21){\special{em: lineto}}
\put(200.00,172.00){\special{em: lineto}}
\put(201.00,172.78){\special{em: lineto}}
\put(202.00,173.56){\special{em: lineto}}
\put(203.00,174.33){\special{em: lineto}}
\put(204.00,175.10){\special{em: lineto}}
\put(205.00,175.86){\special{em: lineto}}
\put(206.00,176.61){\special{em: lineto}}
\put(207.00,177.36){\special{em: lineto}}
\put(208.00,178.10){\special{em: lineto}}
\put(209.00,178.84){\special{em: lineto}}
\put(210.00,179.57){\special{em: lineto}}
\put(211.00,180.30){\special{em: lineto}}
\put(212.00,181.02){\special{em: lineto}}
\put(213.00,181.73){\special{em: lineto}}
\put(214.00,182.44){\special{em: lineto}}
\put(215.00,183.14){\special{em: lineto}}
\put(216.00,183.84){\special{em: lineto}}
\put(217.00,184.53){\special{em: lineto}}
\put(218.00,185.22){\special{em: lineto}}
\put(219.00,185.91){\special{em: lineto}}
\put(220.00,186.58){\special{em: lineto}}
\put(221.00,187.26){\special{em: lineto}}
\put(222.00,187.93){\special{em: lineto}}
\put(223.00,188.59){\special{em: lineto}}
\put(224.00,189.25){\special{em: lineto}}
\put(225.00,189.90){\special{em: lineto}}
\put(226.00,190.55){\special{em: lineto}}
\put(227.00,191.20){\special{em: lineto}}
\put(228.00,191.84){\special{em: lineto}}
\put(229.00,192.47){\special{em: lineto}}
\put(230.00,193.10){\special{em: lineto}}
\put(231.00,193.73){\special{em: lineto}}
\put(232.00,194.35){\special{em: lineto}}
\put(233.00,194.97){\special{em: lineto}}
\put(234.00,195.58){\special{em: lineto}}
\put(235.00,196.19){\special{em: lineto}}
\put(236.00,196.80){\special{em: lineto}}
\put(237.00,197.40){\special{em: lineto}}
\put(238.00,198.00){\special{em: lineto}}
\put(239.00,198.59){\special{em: lineto}}
\put(240.00,199.18){\special{em: lineto}}
\put(241.00,199.77){\special{em: lineto}}
\put(242.00,200.35){\special{em: lineto}}
\put(243.00,200.93){\special{em: lineto}}
\put(244.00,201.50){\special{em: lineto}}
\put(245.00,202.07){\special{em: lineto}}
\put(246.00,202.64){\special{em: lineto}}
\put(247.00,203.20){\special{em: lineto}}
\put(248.00,203.76){\special{em: lineto}}
\put(249.00,204.31){\special{em: lineto}}
\put(250.00,204.87){\special{em: lineto}}
\end{picture}

{\small \it
\begin{center}
Figure 10.
\\
The form of  $I_{00}$ as a function of $A$.
   \end{center}
}

\end{figure}
Here the most interesting function $I_{00}= \int_0^{\infty} \exp(-G) dx$
is drawn.
Two curves are shown. the lower curve corresponds to the precise
numerical solution, i.e. to $\int_0^{\infty}
\exp(-G) dx$. The upper curve corresponds to $B(z=\infty) / A = \int_0^z
\exp(-G_2) dx$. This curve is very
close to numerical solution.
It is very important that both curves
resemble  a straight line.
So, already the linear approximation on $A$
corresponding to the account of the first leading (linear) term will be good.
It means that in the last expression one can take
instead of $B(z=\infty)$ the value $A \int_0^{\infty} \exp(-G_0) dx$
(on the base of the first iteration one  puts
$\sim A \int_0^{\infty} f_1 dx$).

\subsection{Wide spectrum of active centers}

Three regions were extracted
in the spectrum of activities: the region of
 active
centers (they are almost exhausted), the region of  centers with
small
activity (they remain practically free during the whole process) and
the intermediate region. In the majority of the
situations
the intermediate region has
 relatively small size in comparison with the
active
region (the passive region has no size - it can be spread to infinity).
Then one can fulfill some further simplifications.

One can transmit the point $w=0$ to the special activity for which
$$
\theta \mid_{w=0} (\infty) = \frac{1}{2} \ \ ,
$$
i.e. the half of the centers became the centers of droplets.
Then the system of the condensation equations
after the rescaling of $w$ can be written as
\begin{equation}
g = a_0 \exp(w) \int_{0}^{z} (z-x)^3 \exp(-G) \theta dx \ \ ,
\end{equation}
\begin{equation}
G = \int_{-\infty}^{w_{+}} g(w)  dw \ \ ,
\end{equation}
\begin{equation}
\theta = \exp( - a_1 \exp(w) \int_{0}^{z} \exp(-G) dx ) \ \ ,
\end{equation}
where $w_{+}$ is the upper boundary of spectrum and $a_0, \ a_1$ are some
constants. Having rescaled $z,x$ one can put $a_0 = 1$. The condition
of
the choice for $w=0$ gives:
\begin{equation}
a_1 = \frac{\ln 2}{\int_{0}^{\infty} \exp(-G) dx} \ \ .
\end{equation}
So, there remains only one parameter $w_{+}$.

Let us explicitly extract the intermediate and the active regions.
One can see that for  two different activities $w_1$ and $w_2$ the
following
equation is valid
\begin{equation}
\frac{\ln(\theta(w_1,x))}{\exp(w_1)}
=
\frac{\ln(\theta(w_2,x))}{\exp(w_2)} \ \ .
\end{equation}
The same is valid for the final values of $\theta$. The value of
$\ln(\theta(w,x)) / \exp(w) $ is invariant for the  different activities
 in one and the same process. So, one can put $w=w_{++} \equiv 2 \div 3$ as the
boundary between the active region and the intermediate region.
In the same manner one can separate the region of the passive centers by
the boundary $w_{--} =  - w_{++}$. One can neglect the substance in the
droplets
on the passive centers and get\footnote{The absence of the active and
intermediate regions ($w_{+} < w_{--}$) means that the condensation occurs
in the pseudo homogeneous way.}
\begin{equation}
G \approx \int_{w_{--}}^{w_{+}} g(w)  dw \ \ .
\end{equation}

One  can denote by $n_{\infty} G_{+}$ the number of the  substance
molecules in the droplets formed on the active centers. The system of
the condensation
equations can be rewritten as
$$
g = a_0 \exp(w) \int_{0}^{z} (z-x)^3 \exp(-G) dx  \ \ ,
$$
$$
G = \int_{-\infty}^{w_{++}} g(w)  dw  + G_{+}  \ \ ,
$$
$$
\theta = \exp( - a_1 \exp(w) \int_{0}^{z} \exp(-G) dx )   \ \ .
$$
Now one can write an evident expression for $G_{+}$
\begin{equation}
G_{+} = \frac{\eta_{*}}{n_{\infty}} (w_{+} - w_{++})   z^3  \ \ .
\end{equation}
The system of the condensation equations now is the following one
$$
g = a_0 \exp(w) \int_{0}^{z} (z-x)^3 \exp(-G) dx \ \ ,
$$
$$
G = \int_{-\infty}^{w_{++}} g(w)  dw  +
\frac{\eta_{*}}{n_{\infty}} (w_{+} - w_{++})   z^3   \ \ ,
$$
$$
\theta = \exp( - a_1 \exp(w) \int_{0}^{z} \exp(-G) dx )  \ \ .
$$
The value of $w_{++}$ is  universal, but the coefficient in the term
${\eta_{*}} (w_{+} - w_{++})   z^3 / n_{\infty} $ depends on parameters.

Having defined
\begin{equation}
G_{-} =\int_{-\infty}^{w_{++}} g(w)  dw   \ \ ,
\end{equation}
one can propose the following estimate
\begin{equation}
G_{-} \leq ( w_{++} -  w_{--} ) \frac{\eta_{*}}{n_{\infty}} z^3  \
\ .
\end{equation}
When $$w_{+} - w_{++} \gg w_{++} -  w_{--}  \sim  4 \div 6 $$
one can approximately get
$$
G = \frac{\eta_{*}}{n_{\infty}} (w_{+} - w_{++})    z^3
\ \ .
$$
Then the first equation isn't necessary at all.
So, there is no necessity to put $a_0$ to $1$ and one can use the arbitrary scale to
cancel another constant. The system looks like
$$
\theta = \exp( - a_1 \exp(w) \int_{0}^{z} \exp(-G) dx )  \ \ ,
$$
$$
G = \frac{\eta_{*}}{n_{\infty}} (w_{+} - w_{++}) z^3  \ \ .
$$
Having rescaled $x, z$, one can put $ {\eta_{*}} (w_{+}
-
w_{++})
= n_{\infty} $ and the system looks like the universal expression:
$$
\theta = \exp( -  \frac{\ln 2}{\int_{0}^{\infty} \exp(-z^3) dx}
\exp(w) \int_{0}^{z} \exp(-z^3) dx )  \ \ .
$$
The behavior of $\theta$ as a function of $z$ for the different values of
$w$ is shown in  Fig. 11.

\begin{figure}



{\small \it
\begin{center}
Figure 11.
\\
Behavior of  $\theta$ as a function of $x$ for different $w$.
   \end{center}
}

\end{figure}

One has to stress that  in the general case of quasi continuous
spectrum all results concerning  the
intermediate approximate solution can be used. One has only to substitute
the sum over  all sorts of centers by the integral over activities
with appropriate normalizing coefficient:
$
\sum_i \rightarrow  \sim \int dw
$.

Moreover, the procedure of intermediate approximate solution
can be directly applied to the situation with
a
mixed quasi continuous spectrum of activities and some discrete $\delta$-like
peaks for several  sorts of heterogeneous centers.

\vspace{2cm}

This remark completes the description of the nucleation  for
the  decay on heterogeneous centers with different activities.
Unfortunately there are no experimental investigations which can
be compared with the theory. It is rather natural because earlier
there were no formulas which could give any physical information
and  experimental investigations would give
only some chaotic experimental data. Now one can easily check
this theory. This publication can be regarded as an invitation
to open experimental investigations in this field. One can use the
same devices as for the ordinary homogeneous nucleation
investigations and calculate the number of droplets.
Experiments on this theory can give information about the
properties of heterogeneous centers in complex systems. The
special interest is to solve some extreme problems to produce
mainly the droplets on a given sort of centers when there are
several  sorts of centers  in the system. With the help
of this theory this prolem can be solved.

\end{document}